\documentclass[11pt]{article}

\usepackage[preprint]{acl}

\usepackage{times}
\usepackage{latexsym}

\usepackage[T1]{fontenc}

\usepackage[utf8]{inputenc}

\usepackage{microtype}

\usepackage{inconsolata}

\usepackage{graphicx}

\usepackage{xcolor,colortbl,caption}

\usepackage[most]{tcolorbox}
\usepackage{comment}
\usepackage{amsmath}
\usepackage{pifont}
\usepackage{tabularx}
\usepackage{caption, booktabs}
\usepackage{algorithm}
\usepackage{algpseudocode}
\usepackage{amssymb}
\usepackage{setspace}
\usepackage{bbm}
\usepackage{arydshln}

\makeatletter
\newcommand\footnoteref[1]{\protected@xdef\@thefnmark{\ref{#1}}\@footnotemark}
\makeatother
\newcolumntype{P}[1]{>{\centering\arraybackslash}p{#1}}

\usepackage{xspace}
\newcommand{\ours}[0]{\textsc{R$^3$-SQL}\xspace}
\newcommand{\oursrm}[0]{\textsc{R$^3$}\xspace}
\newcommand{\oursprm}[0]{\textsc{R$^3$-Point-32B}\xspace}
\newcommand{\ourslrm}[0]{\oursrm-7B\xspace}
\usepackage{graphicx}
\usepackage{adjustbox}
\makeatletter
\newcommand{\thickhline}{
    \noalign {\ifnum 0=`}\fi \hrule height 1pt
    \futurelet \reserved@a \@xhline
}
\newcolumntype{"}{@{\hskip\tabcolsep\vrule width 1pt\hskip\tabcolsep}}
\makeatother
\usepackage{mathabx}
\usepackage{amsfonts}

\makeatletter
\newcommand*{\blackleq}{
  \mathrel{
    \mathpalette\@blackleq{}
  }
}
\newcommand*{\@blackleq}[2]{
  \vcenter{
    \m@th
    \setbox0=\hbox{$#1\mkern3mu$}
    \setbox2=\hbox{$#1\vcenter{}$}
    \setbox4=\hbox{\raisebox{-\ht2}[.2pt][.2pt]{$#1-$}}
    \hbox{$#1\blacktriangleleft$}
    \nointerlineskip
    \kern\wd0 
    \copy4 
  }
}
\makeatother
\usepackage{dsfont}
\usepackage{multirow}
\usepackage{kotex} 
\newcommand{\edit}[1]{\normalfont #1}

\usepackage{mathtools}

\definecolor{my_blue}{RGB}{0,112,192}

\usepackage{placeins}

\usepackage{tikz}

\usepackage{minted}
\usepackage{fvextra}

\DefineVerbatimEnvironment{WideMinted}{Verbatim}
{breaklines, fontsize=\small, frame=lines, linenos, breakanywhere, xleftmargin=0pt, xrightmargin=0pt, width=\textwidth}

\usepackage{stackengine}

\makeatletter
\def\th@definition{%
  \normalfont 
  \thm@headpunct{.}
}
\makeatother

\usepackage{enumitem}
\usepackage{wrapfig}
\usepackage{subcaption}

\usepackage{makecell}
\usepackage{cuted}
\usepackage{tablefootnote}
\renewcommand{\dbltopfraction}{0.9}
\title{\edit{\textbf{\ours: Ranking Reward and Resampling for Text-to-SQL}}}

\author{
Hojae Han$^{1}$\thanks{Equal contribution.} Yeonseok Jeong$^{2}$\footnotemark[1]\thanks{Work done while at Snowflake.} Seung-won Hwang$^{2}$\thanks{Corresponding author.} $ $ Zhewei Yao$^{3}$ Yuxiong He$^{3}$\\
$^{1}$ETRI, $^{2}$Seoul National University, $^3$Snowflake AI Research \\
\texttt{hojae.han@etri.re.kr} \texttt{\{jys3136,seungwonh\}@snu.ac.kr}\\
\texttt{\{zhewei.yao,yuxiong.he\}@snowflake.com} \\
} 

\begin{document}
\maketitle

\begin{abstract}
Modern Text-to-SQL systems generate multiple candidate SQL queries and rank them to judge a final prediction. However, existing methods face two limitations. First, they often score functionally equivalent SQL queries inconsistently despite identical execution results. Second, ranking cannot recover when the correct SQL is absent from the candidate pool. We propose \ours, a Text-to-SQL framework that addresses both issues through unified reward for ranking and resampling. \ours first groups candidates by execution result and ranks groups for consistency. To score each group, it combines a pairwise preference across groups with a pointwise utility from the best group rank and size, capturing relative preference, consistency, and candidate quality. To improve candidate recall, \ours introduces agentic resampling, which judges the generated candidate pool and selectively resamples when the correct SQL is likely absent. \ours achieves 75.03 execution accuracy on BIRD-dev, a new state of the art among methods using models with disclosed sizes, with consistent gains across five benchmarks.
\end{abstract}
\section{Introduction}
\label{sec:intro}

\begin{figure*}[t]
    \centering
    \includegraphics[width=1.0\linewidth]{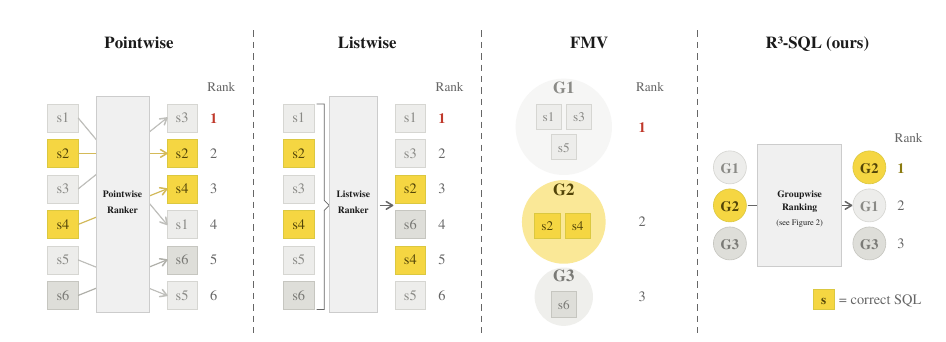}
    \caption{Comparison of ranking strategies for six SQL candidates, where $s_2$ and $s_4$ (yellow) are correct. Both pointwise and listwise rankers rank an incorrect candidate first. Functional majority voting (FMV) groups candidates by execution result but ranks the largest (incorrect) group $G_1$ first. \ours re-ranks groups using pointwise and listwise signals with group size, correctly placing $G_2$ at rank 1.}
    \label{fig:main}
\end{figure*}

With advances in large language models (LLMs), Text-to-SQL systems adopt a generate-then-rank paradigm: an LLM samples multiple candidate SQL queries, then a ranker selects the best candidate from the pool~\cite{pourreza2025chasesql,agrawal2025text2sql,XiYanSQL} (Figure~\ref{fig:main}). Existing rankers can be categorized into \textit{pointwise}~\citep{agrawal2025text2sql}, which scores each candidate $s_i$ independently, or \textit{listwise}
comparing multiple candidates $\{s_1,\dots,s_n\}$ jointly as Figure~\ref{fig:main} illustrates. 

However, both rankers face two limitations.
First, rankers assign inconsistent scores to functionally equivalent SQL queries that differ in surface form but produce identical execution results~\cite{launer2026majority,LI2026129523}. 
For example, $s_2$ and $s_4$ in Figure~\ref{fig:main} produce the same execution result but receive different scores in both pointwise and listwise rankers~\cite{long2025precise}. We term this \textit{functional inconsistency}. 

A common remedy is to group functionally equivalent candidates before ranking~\citep{sheng2025csc,launer2026majority,LI2026129523}. However, existing methods rank groups by size alone (functional majority voting in Figure~\ref{fig:main}), so a small but correct group can still be outranked by a larger incorrect one.
Second, rankers assume that the correct SQL already exists in the candidate pool. When the generator fails to produce a correct candidate, no ranking strategy can recover the correct answer. This limitation is well-known as \textit{bounded recall} in the information retrieval literature~\cite{wang2011cascade,rathee2025guiding,macavaney2022adaptive,zhang2021learning}, addressed by resampling, yet remains unexplored in Text-to-SQL.

We propose \ours, a Text-to-SQL framework that addresses both limitations through unified  modeling for \underline{r}eward \underline{r}anking and \underline{r}esampling.
First, to mitigate functional inconsistency, \ours groups candidates by execution result and ranks groups rather than individual SQL queries. 
Figure~\ref{fig:main} illustrates 
our distinction, groupwise ranking combining two complementary signals. The first is listwise comparisons of pairs across groups, capturing relative preference. The second is a groupwise utility signal derived from both the rank of the representative sample and its size. Figure~\ref{fig:overview} illustrates how we unify three signals: relative inter-group preference, inner-group consistency, and individual candidate quality, and agentic resampling. 
Second, 
resampling ensures the candidate pool includes a correct generation, by predicting
if  a correct SQL query in the pool and resampling otherwise. 

Experiments on five Text-to-SQL benchmarks show that \ours consistently improves execution accuracy over prior ranking-based approaches. On BIRD-dev~\cite{liCanLLMAlready2023}, \ours achieves 75.03 execution accuracy, establishing a new state of the art among methods using models with disclosed sizes (Table~\ref{tab:cross_benchmark}). Further analysis confirms that each component contributes: groupwise scoring eliminates score variance among equivalent SQLs, the consistency objective improves input-order robustness by +11.89 pp, and agentic resampling raises candidate recall by +3.92 pp.

Our contributions are as follows:
\begin{itemize}[leftmargin=*,nosep]
\item We identify two core limitations of ranking-based Text-to-SQL pipelines: functional inconsistency and bounded recall.
\item We propose a group-based ranking framework that combines pointwise and listwise reward signals to resolve functional inconsistency.
\item We introduce agentic resampling to selectively expand the candidate pool when the correct SQL is likely absent.
\item We demonstrate state-of-the-art performance on five Text-to-SQL benchmarks and provide analysis of each component's contribution.
\end{itemize}

\section{Related Work}
\label{sec:related_work}

Recent Text-to-SQL systems follow a generate-then-rank pipeline: sample multiple candidate SQL programs, then rank or filter to select a final prediction. As discussed in \S\ref{sec:intro}, two key challenges shape performance: functional inconsistency and bounded recall. Existing systems vary in how they address these challenges, as summarized in Table~\ref{tab:contribution}.

\begin{table*}[t]
\centering
\small
\setlength{\tabcolsep}{6pt}
\resizebox{\textwidth}{!}{
\begin{tabular}{l c c c c c}
\toprule
Method & LLM Size & Grouping & Ranker & Resample when & EX (\%) \\
\midrule
\multicolumn{6}{c}{\textit{Model size undisclosed (UNK)}} \\
\midrule
Contextual\text{-}SQL        & UNK & --         & pointwise              & never                    & 73.50 \\
CHASE\text{-}SQL             & UNK & --         & listwise               & syntax errors            & 74.90 \\
Agentar\text{-}Scale\text{-}SQL & UNK & --      & listwise               & syntax \& semantic errors & 74.90 \\
XiYan\text{-}SQL             & UNK & \checkmark & listwise + FMV          & syntax errors            & 73.34 \\
OpenSearch\text{-}SQL        & UNK & \checkmark & FMV                     & syntax errors            & 69.30 \\
\midrule
\multicolumn{6}{c}{\textit{Model size disclosed}} \\
\midrule
\textbf{\ours}               & \textbf{32B} & \textbf{\checkmark} & \textbf{groupwise (point + list) + FMV} & \textbf{syntax \& semantic errors} & \textbf{75.03} \\
CSC\text{-}SQL               & 70B  & \checkmark & FMV                     & always                   & 71.69 \\
CSC\text{-}SQL               & 32B  & \checkmark & FMV                     & always                   & 71.33 \\
Agentar\text{-}Scale\text{-}SQL & 32B & --       & listwise               & syntax \& semantic errors & 71.12 \\
XiYan\text{-}SQL             & 32B  & \checkmark & listwise + FMV          & syntax errors            & 67.01 \\
\bottomrule
\end{tabular}}
\caption{Comparison of recent Text-to-SQL systems on BIRD-dev.
\emph{Grouping} indicates whether candidates are grouped by execution result.
\emph{Ranker} specifies the scoring method; FMV denotes functional majority voting by group size.
\emph{Resample when} denotes the condition that triggers candidate resampling or refinement.
}
\label{tab:contribution}
\end{table*}

\textbf{Contextual\text{-}SQL}~\citep{agrawal2025text2sql} trains a pointwise ranker to score each candidate independently without grouping, so functional inconsistency persists. It conducts extensive initial sampling but does not resample or refine candidates.
\textbf{CHASE\text{-}SQL}~\citep{pourreza2025chasesql} employs a pairwise ranker (listwise with $n{=}2$) that compares candidates relatively, but without execution-based grouping, functional inconsistency persists. It refines only syntax errors, so bounded recall from semantic errors remains unresolved.
\textbf{Agentar\text{-}Scale\text{-}SQL}~\citep{wang2025agentar} employs a listwise ranker without grouping, and enhances recall by refining candidates through both syntax and semantic checks via an agent.
\textbf{XiYan\text{-}SQL}~\citep{XiYanSQL} ensembles multiple generators and groups candidates by execution result, but ranks groups by size, so a small but correct group can be outranked by a larger incorrect one. It refines only syntax errors, so bounded recall from semantic errors remains unresolved.
\textbf{OpenSearch\text{-}SQL}~\citep{xie2025opensearch} adopts groupwise voting for selection, but this heuristic is limited compared to learned rankers and similarly relies on group size. It refines only syntax errors.
\textbf{CSC\text{-}SQL}~\citep{sheng2025csc} groups candidates by execution result and applies merge-revision over top-2 groups. However, the revision is applied indiscriminately even when a correct candidate already exists, which can introduce unnecessary errors. The absence of learned rankers limits precision in the final selection.

\paragraph{{Our Distinction.}}
As shown in Table~\ref{tab:contribution}, \ours addresses both limitations. For functional inconsistency, \ours groups candidates by execution result and ranks groups using both pointwise and listwise signals rather than group size alone. For bounded recall, \ours introduces agentic resampling that judges whether the candidate pool likely contains a correct SQL and selectively resamples when it does not. This achieves the highest EX on BIRD-dev among systems with disclosed model sizes.
\section{{\ours}}

As illustrated in Figure~\ref{fig:overview}, \ours operates in two phases: exploration and exploitation. We first describe how candidates are grouped and ranked to resolve functional inconsistency (\S\ref{sec:improving_pointwise}), then address bounded recall via agentic resampling (\S\ref{sec:generation_bias_mitigation}), followed by position-bias mitigation in the listwise ranker (\S\ref{sec:improving_listwise}). We summarize the overall framework in \S\ref{sec:overall}.

\subsection{Improving Functional Inconsistency}
\label{sec:improving_pointwise}
Functional inconsistency arises when functionally equivalent SQL queries receive inconsistent scores due to superficial differences (e.g., token order or stylistic variations). \ours\ addresses this by evaluating candidates at the level of \emph{execution semantics} rather than individually. 

\paragraph{{Groupwise Scoring.}}
{Instead of scoring each candidate independently, we group SQL candidates that yield the same execution result into a single cluster. Each group thus represents one distinct semantic outcome. Formally, let $\mathcal{G}=\{g_1,\dots,g_M\}$ be the set of groups after executing all candidates and grouping those with identical results. By aggregating scores over execution-equivalent groups, we neutralize the noise introduced by superficial textual differences.
} 

Grouping ensures consistent treatment within each group. The remaining question is how to rank across groups. We combine two signals: a cross-group preference from pairwise comparisons, which leverages multiple observations per group pair, and a utility from pointwise scoring and group size. The cross-group signal serves as the primary ranking criterion; when its margin is too small to be reliable, the pointwise utility breaks ties.

\paragraph{{Cross-Group Preference Signal.}}
We utilize a \emph{pairwise}\footnote{We use pairwise as a special case of listwise ranking with $n{=}2$.} ranker that compares candidates from different groups and produces a relative preference signal. 
Specifically, we train a dedicated pairwise ranker that takes the database schema $db$, the natural language question $x$, and two SQL candidates (along with their execution results) as input, and outputs a preference indicating which candidate is more likely to be correct.

Motivated by the Bradley--Terry model~\cite{bradley1952rank}, which views pairwise preferences as noisy observations of latent quality scores via
\begin{equation}
\label{eq:margin}
P(g_i \succ g_j) = \sigma(r_i - r_j),
\end{equation}
where $\sigma$ is the logistic sigmoid and $r_i$, $r_j$ are latent utilities,
we estimate the group-level preference $P(g_i \succ g_j)$ by aggregating pairwise comparisons between all candidates $s_i \in g_i$ and $s_j \in g_j$, i.e.,
$\frac{1}{|g_i|\,|g_j|}\sum_{s_i\in g_i}\sum_{s_j\in g_j} v(s_i,s_j)$,
where $v(s_i,s_j)\in\{0,1\}$ is the ranker's vote preferring $s_i$ over $s_j$ ($1$ if $s_i$ is judged better, $0$ otherwise). To avoid noise in close or ambiguous comparisons (e.g. if both candidates are incorrect), we apply a threshold $\tau$ to discount uncertain judgments.\footnote{We fix $\tau{=}0.05$ across the settings (see Appendix~\ref{appendix:tau}).} Only a sufficiently large margin counts as a decisive preference of $g_i$:
\begin{equation}
\label{eq:threshold}
\tilde P(g_i \succ g_j) \;=\; 
\begin{cases}
+1, & \text{if } P(g_i \succ g_j)\ge \tau,\\
0, & \text{otherwise,}
\end{cases}
\end{equation}
and we define the pairwise group score as the number of decisive wins against all other groups:

\begin{equation}
r_{\text{list}}(g_i)\;=\;\sum_{j \neq i}\tilde P(g_i \succ g_j)\,. 
\end{equation}
This pairwise procedure produces a group ordering that is robust to functional inconsistency. By comparing candidates side-by-side, our ranker can consistently select correct SQLs above those with only superficial token advantages.

\begin{figure*}[t]
    \centering
    \includegraphics[width=0.85\linewidth]{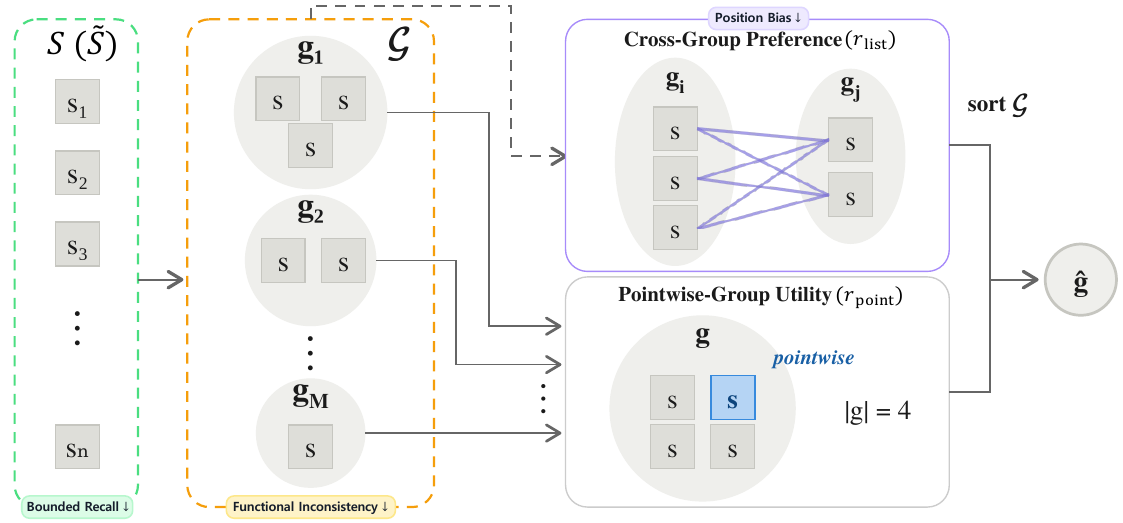}
    \caption{Groupwise Ranking by \ours.}
    \label{fig:overview}
\end{figure*}

\paragraph{{Pointwise-Group Utility Signal.}}
As complementary signal for group utility,
we also use the \emph{pointwise ranker} that provides scores independent of input order.
We also incorporate a confidence measure function $w(g)$.
For each execution-equivalent group $g$, we define
\begin{align}
\label{eq:anchor}
r_{\text{point}}(g) = w(g)\cdot u(g) 
\end{align}
where $w(g){=}|g|$ to reflect execution-level self-consistency and further stabilize rankings. We set $u(g){=}\max_{s\in g}{RR}_s$, preserving the strongest evidence within each group, denoted as a colored rectangle in group $g$ in Figure~\ref{fig:overview}. This follows representation fusion literature~\cite{10.1145/584792.584854}, where $RR_s$ denotes the reciprocal rank of candidate $s$ according to the pointwise ranker. 
As $r_{\text{point}}(g)$ is independent of input position, it
overcomes positional effects from $r_{\text{list}}$.

\subsection{\textbf{Mitigating the Bounded Recall Problem}}
\label{sec:generation_bias_mitigation}
As introduced in \S\ref{sec:intro}, even the best ranking strategy cannot succeed if the correct SQL is absent from the candidate pool.
To address this, \ours\ augments the candidate sampling stage with an LLM-based agent that diagnoses coverage gaps and triggers targeted resampling when necessary, so the ranker operates on a pool more likely to contain a correct SQL.

\paragraph{\textbf{Agentic Resampling.}}
Our candidate sampling procedure is a two-pass process: initial pool generation and selective resampling.
First, given a database schema $db$ and natural language question $x$, we prompt a base LLM 
to produce the initial pool $S=\{s_1,\dots,s_n\}$ of $n$ SQL candidates. 
Then, an LLM agent $f$ examines whether $S$ contains the correct SQL, denoted as $f(S) \in \{0, 1\}$. 
When the agent $f$ decides $S$ is insufficient, i.e., $f(S){=}0$, it discards $S$ and triggers the resampling: 
\begin{align}    
\label{eq:pool_selection}
S =
\begin{cases}
\tilde{S}, & \text{if } f(S)=0 ,\\[4pt]
S, & \text{otherwise,}
\end{cases}
\end{align}
where $\tilde{S}$ is the resampled pool. We sample a larger set $\tilde{S}=\{\tilde{s}_1,\dots,\tilde{s}_m\}$ of $m$ ($m > n$) new candidate SQLs from the base LLM. 
To focus on the most promising of these resampled candidates, we use the pointwise ranker (as used for anchoring) to score each $\tilde{s}_i$. 
The top-$n$ scored queries are then selected to form the resampled pool, denoted simply as $\tilde{S}$ for brevity, matching the original pool size.

\subsection{{Improving Listwise Ranking}}
\label{sec:improving_listwise}
While the above ranking addresses functional inconsistency at the group level, the listwise ranker itself can be sensitive to the input order of candidates, known as position bias.
\ours\ counters this by training the listwise ranker with a position-consistency objective.

\paragraph{{Position-Consistency Objective.}}
\label{sec:pairwise_rm}
We train a listwise ranker to be robust to input order by presenting each correct–incorrect SQL pair $(s^+, s^-)$ in both original and swapped positions.
Within the GRPO framework~\citep{shao2024deepseekmath}, we introduce a consistency reward, inspired by the position-invariance objective of J1~\citep{whitehouse2025j1}, to reinforce preferences that remain correct under both orderings.
Specifically, we decompose the reward into a base term and an auxiliary consistency term:
\begin{align}
\label{eq:consistency_reward}
R = R_{\mathrm{base}} + \lambda_{{c}} \, R_{{c}},
\end{align}
where $R_{\mathrm{base}} \in \{0,1\}$ indicates whether the preference decision is correct, and $R_{{c}} \in \{0,1\}$ indicates whether the correct decision is preserved under order perturbation.
We set $\lambda_{{c}}{=}0.5$ without additional tuning, following common RL practice of assigning a modest weight to auxiliary objectives~\cite{brockman2016openai,schulman2017proximal}.
As a result, each correct--incorrect pair receives reward $0$ if the decision is incorrect, $1$ if it is correct, and $1.5$ if it is correct and remains consistent across both input orders.

\subsection{{Overall Framework}}
\label{sec:overall}
As illustrated in Figure~\ref{fig:overview}, \ours\ operates as an exploration-exploitation framework: the first phase expands the candidate pool via agentic resampling to address bounded recall, while the second phase ranks candidates through group-based ranking to resolve functional inconsistency.

\paragraph{\textbf{Agentic resampling for Recall.}}
A base LLM generates an initial pool $S$ of candidate SQLs.
An LLM agent $f$ audits $S$, deciding to retain or replace it with a resampled pool $\tilde S$ according to the selection rule in Eq.~\eqref{eq:pool_selection}.

\paragraph{\textbf{Pointwise-listwise ranking for Precision.}}
Execute all candidates in $S$ (or $\tilde S$) and group those with identical execution results into $\mathcal{G}=\{g_1,\dots,g_M\}$, each representing a distinct semantic outcome.
Compute listwise preferences using Eqs.~\eqref{eq:margin}–\eqref{eq:threshold},
and combine them with the pointwise group utility $r_{\text{point}}(g)$ from Eq.~\eqref{eq:anchor}. Groups are then sorted by the lexicographic ordering:
\begin{align}
\label{eq:lexicographic}
\hat{\mathcal{G}} = \mathrm{sort}(\mathcal{G}) \;\; \text{by} \;\; \big( r_{\text{list}}(g),\; r_{\text{point}}(g) \big),
\end{align}
where tuples are compared left-to-right. Because the threshold $\tau$ in Eq.~\eqref{eq:threshold} discards preference margins too small to distinguish from noise, groups with similar listwise quality receive identical $r_{\text{list}}$ scores. For such ties, $r_{\text{point}}$ provides a fallback based on absolute candidate quality and group consensus.
Since the lexicographic ordering operates on group-level signals, we perform one final comparison at the individual SQL level. Let $\{g',g''\} = \mathrm{Top2}(\hat{\mathcal{G}})$; the final group is selected as:
\begin{align}
\label{eq:final}
\hat{g} = \begin{cases} g', & \text{if } P(g' \succ g'') > 1/2,\\ g'', & \text{otherwise,} \end{cases}
\end{align}
and the candidate with the highest pointwise rank within $\hat{g}$ is returned as the final prediction.
\section{Experiments}

\begin{table*}[t]
    \centering
    \small
    \resizebox{\textwidth}{!}{
    \begin{tabular}{l l c c c c c c}
        \toprule
        \multirow{2}{*}{SQL Selection Method} & 
        \multirow{2}{*}{Ranker} & 
        \multirow{2}{*}{BIRD-dev} & 
        \multirow{2}{*}{Spider-test} & 
        \multirow{2}{*}{Spider-DK} & 
        \multirow{2}{*}{EHRSQL} & 
        \multirow{2}{*}{\shortstack{Science\\Benchmark}} &
        \multirow{2}{*}{\textbf{Avg.}} \\ 
        \\
        \midrule
        CSC-SQL & FMV & 71.58 & 86.64 & 76.97 & 41.04 & 56.68 & 66.58 \\
        Contextual-SQL & Pointwise & 73.14 & 86.36 & 75.50 & 41.41 & 63.13 & 67.91 \\
        CHASE-SQL & Listwise & 73.34 & 86.18 & 75.94 & 44.44 & 63.59 & 68.70 \\
        XiYan-SQL & Listwise + FMV & 72.03 & 85.89 & 75.28 & 43.43 & 63.59 & 68.04 \\
        \ours & Groupwise (Point + List) + FMV & \textbf{75.03} & \textbf{87.19} & \textbf{77.92} & \textbf{46.30} & \textbf{66.82} & \textbf{70.65} \\
        \bottomrule
    \end{tabular}}
    \normalsize
    \caption{EX comparison of different selection methods across five benchmarks, where SQL candidates were generated by Arctic-Text2SQL-R1-32B ($T{=}0.8$).
    The same pointwise (\oursprm) and listwise (\ourslrm; \S\ref{sec:pairwise_rm}) rankers were consistently used across the methods.}
    \label{tab:cross_benchmark}
\end{table*}
\subsection{Experimental Setup}
Appendix~\ref{appendix:details} explains the implementation details.
\paragraph{Benchmarks.}
We evaluate \ours on 5 widely-used Text-to-SQL benchmarks:
BIRD~\citep{liCanLLMAlready2023}, 
Spider~\citep{yuSpiderLargeScaleHumanLabeled2018}, 
Spider-DK~\citep{gan-etal-2021-exploring}, 
EHR-SQL~\citep{10.5555/3600270.3601404}, and
ScienceBenchmark~\citep{zhang2023sciencebenchmark}. 
Unlike the first three cross-domain benchmarks, EHRSQL and ScienceBenchmark are domain-specific, testing out-of-domain generalization capabilities.
Detailed statistics of these benchmarks are shown in Appendix Table~\ref{tab:benchmark_statistics}.

\paragraph{Metrics.}
Throughout the experiments, we use execution accuracy (EX) as the main metric, which is the official evaluation metric for the benchmarks.
EX evaluates whether the execution result of the predicted SQL matches that of the ground-truth SQL when executed on SQLite.
For fair evaluation, we exclude questions where the ground-truth SQL produces empty execution result due to issues such as database size limitation or timeout error.

\paragraph{Baselines.}
We use CSC-SQL~\citep{sheng2025csc}, Contextual-SQL~\citep{agrawal2025text2sql}, CHASE-SQL~\citep{pourreza2025chasesql}, and XiYan-SQL~\citep{XiYanSQL} as our baselines.
To ensure fair comparison of selection algorithm, we employ the same pointwise ranker (\oursprm) and listwise ranker (\ourslrm; \S\ref{sec:pairwise_rm}) across all baseline methods.\footnote{Except for Contextual-SQL, their original rankers are unreleased.}
\oursprm\ is trained from the pointwise ranker used in Contextual-SQL (Contextual-RM-32B) on BIRD-train; details in Appendix~\ref{appendix:details}.

\subsection{Experimental Results}
Table~\ref{tab:cross_benchmark} presents the comparative results across five diverse benchmarks. \ours consistently outperforms all baselines, achieving the highest EX on every dataset. \ours is the only method to break the 70\% ceiling, achieving an average EX of 70.65\%.
\begin{table}[t]
    \centering
    \small
    \begin{tabular}{l c c}
        \toprule
        \multirow{2}{*}{Method} & \multirow{2}{*}{\shortstack{Score Variance \\ (z-score; $\downarrow$)}} & \multirow{2}{*}{EX (\%)} \\
        \\
        \midrule
        Contextual-SQL & 0.8571 & 73.14 \\
        \ours\ \textit{w/o} \ourslrm & \textbf{0.0000} & 73.47 \\
        \ours & \textbf{0.0000} & \textbf{75.03} \\
        \bottomrule
    \end{tabular}
    \normalsize
    \caption{Functional inconsistency mitigation by \ours on BIRD-dev. Score variance is measured among SQL candidates that yield identical execution results (lower is better).}
    \label{tab:token_bias_variance}
\end{table}
\begin{figure}[t]
    \centering
    \includegraphics[width=1.0\linewidth]{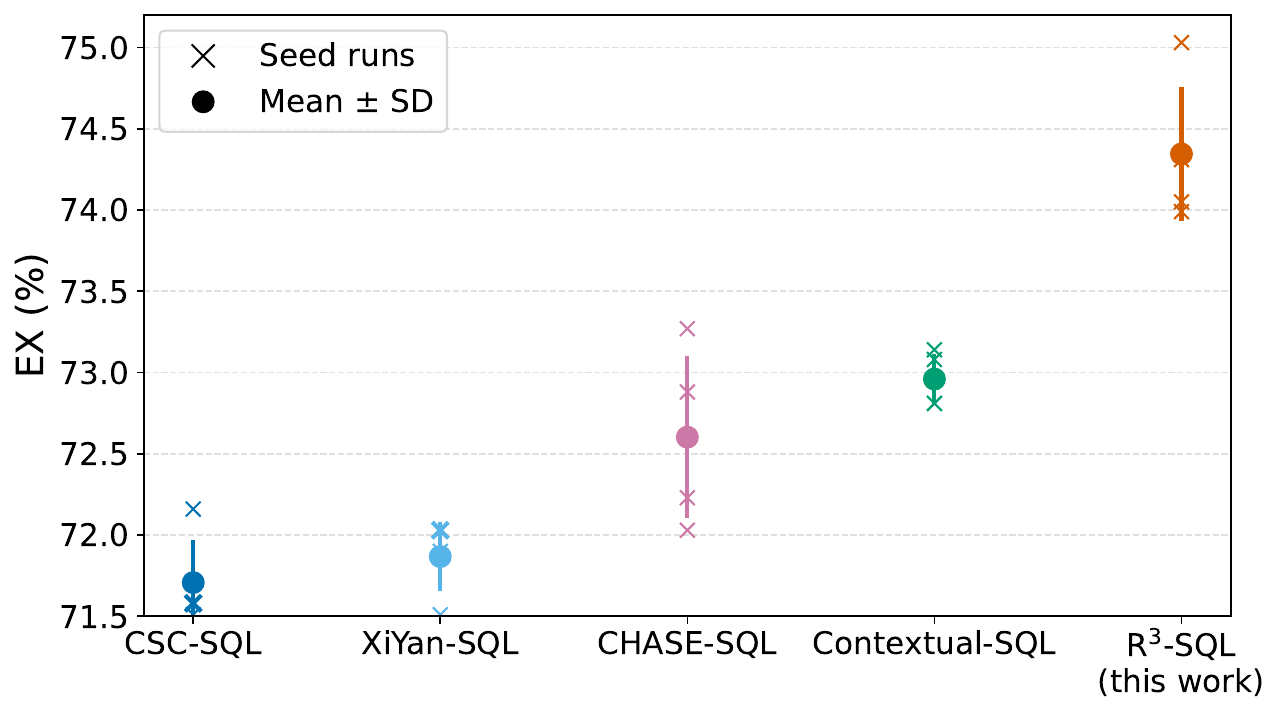}
    \caption{EX stability across 4 random seeds on BIRD-dev. Crosses indicate individual runs, and circles with error bars represent the mean $\pm$ SD. Candidates were generated by Arctic-Text2SQL-R1-32B ($T{=}0.8$).}
    \label{fig:seed-acc-bird}
\end{figure}

\section{Analysis}
\subsection{\textbf{Functional Inconsistency Mitigation}}
\label{analysis:token_bias}
We define functional inconsistency as occurring when functionally equivalent SQLs with identical execution results but different surface forms receive (i) inconsistent scores, and (ii) one such incorrect variant is ranked above the correct SQL. In our evaluation, (i) is measured by the score variance among SQLs with the same execution results, and (ii) is reflected in EX. As shown in Table~\ref{tab:token_bias_variance}, Contextual-SQL solely using a pointwise ranker produces non-zero score variance. \ours's groupwise scoring assigns a single score per execution result, reducing this variance to 0.0 by design. Adding the listwise ranker \ourslrm raises EX by +1.56 pp via better aligning group scores with correctness.

\begin{table*}[t]
    \centering
    \small
    \setlength{\tabcolsep}{6pt}
    \begin{tabular}{l c c c c c c}
        \toprule
        \multirow{2}{*}{Group Scoring} &
        \multirow{2}{*}{BIRD-dev} &
        \multirow{2}{*}{Spider-test} &
        \multirow{2}{*}{Spider-DK} &
        \multirow{2}{*}{EHRSQL} &
        \multirow{2}{*}{\shortstack{Science\\Benchmark}} &
        \multirow{2}{*}{\textbf{Avg.}} \\
        \\
        \midrule
        Pointwise & 73.14 & 86.36 & 75.50 & 41.41 & 63.13 & 67.91 \\
        Pointwise (avg.) & 71.90 & 86.03 & 74.39 & 39.73 & 62.21 & 66.85 \\
        Listwise & 73.34 & 86.66 & 76.82 & \textbf{46.63} & \textbf{67.28} & 70.15 \\
        \ours & \textbf{75.03} & \textbf{87.19} & \textbf{77.92} & 46.30 & 66.82 & \textbf{70.65} \\
        \bottomrule
    \end{tabular}
    \normalsize
    \caption{Comparison of group scoring strategies across five benchmarks. \textit{Pointwise} scores each candidate independently without grouping. \textit{Pointwise (avg.)} ranks groups by the average pointwise score of their members. \textit{Listwise} uses only the pairwise preference signal $r_{\text{list}}$. \ours combines $r_{\text{list}}$ and $r_{\text{point}}$ (max-based) via lexicographic ordering (Eq.~\ref{eq:lexicographic}).}
    \label{tab:group_scoring}
\end{table*}

\begin{table*}[t]
    \centering
    \small
    \setlength{\tabcolsep}{6pt}
    \begin{tabular}{l c c c c c c}
        \toprule
        \multirow{2}{*}{SQL Selection Method} & 
        \multirow{2}{*}{BIRD-dev} & 
        \multirow{2}{*}{Spider-test} & 
        \multirow{2}{*}{Spider-DK} & 
        \multirow{2}{*}{EHRSQL} & 
        \multirow{2}{*}{\shortstack{Science\\Benchmark}} &
        \multirow{2}{*}{\textbf{Avg.}} \\
        \\
        \midrule
        \ours  & \textbf{84.62} & \textbf{93.57} & \textbf{88.30} & \textbf{67.85} & \textbf{79.26} & \textbf{82.72} \\
        \phantom{0}\textit{w/o agentic resampling}  & 81.23 & 92.21 & 87.20 & 58.25 & 75.11 & 78.80 \\
        \bottomrule
    \end{tabular}
    \normalsize
    \caption{
        Bounded recall mitigation by \ours's agentic resampling. Values represent the recall of generated candidates, which corresponds to the ranking upper bound (i.e., the maximum achievable EX by an optimal ranker).
    }
    \label{tab:bounded_recall}
\end{table*}

\begin{table}[t]
    \centering
    \small
    \begin{tabular}{lc}
        \toprule
        Method & 
        Input Consistency (\%;$\uparrow$) \\
        \midrule
        \ourslrm & \textbf{57.49} \\
        \phantom{0}\textit{w/o consistency reward} & 45.60 \\
        \phantom{00}\textit{w/o GRPO} & 37.82 \\
        \bottomrule
    \end{tabular}
    \normalsize
    \caption{Position bias mitigation in \ours's listwise ranker via consistency reward during GRPO training. We evaluate consistency across swapped candidate orders (pos-neg vs. neg-pos) on BIRD-dev using candidates from Arctic-Text2SQL-R1-32B ($T{=}0.8$)}
    \label{tab:consistency_reward}
\end{table}

\subsection{Reproducibility across Seeds}
To assess robustness against sampling randomness, we performed four independent runs on BIRD-dev following the setup in Table~\ref{tab:cross_benchmark}. As illustrated in Figure~\ref{fig:seed-acc-bird}, while the relative rankings of baselines fluctuate due to variations in candidate pools, \ours consistently maintains the lead across all seeds. Most notably, the lowest performance recorded for \ours (73.99) surpasses the absolute peak performance of the strongest baseline (CHASE-SQL: 73.27), demonstrating that our method ensures superior performance regardless of generation noise.

\subsection{\textbf{Group Scoring Strategy}}
\label{analysis:group_scoring}
Table~\ref{tab:group_scoring} compares group scoring strategies. Among single-signal approaches, Listwise performs best by leveraging multiple observations per group pair, followed by Pointwise (max-based). Averaging pointwise scores within a group (Pointwise avg.) performs worst, as weaker members dilute the strongest candidate's signal. \ours combines $r_{\text{list}}$ and $r_{\text{point}}$ via lexicographic ordering and achieves the highest average EX (70.65). On in-domain benchmarks, \ours outperforms Listwise by +1.11 pp on average. On out-of-domain benchmarks, Listwise slightly outperforms \ours, as the domain-limited pointwise ranker introduces noise in unseen domains (see Appendix~\ref{appendix:domain_gen}, Table~\ref{tab:indomain_ood}).

\subsection{\textbf{Bounded Recall Mitigation}}
\label{analysis:bounded_recall}
Table~\ref{tab:bounded_recall} demonstrates the effectiveness of \ours's agentic resampling in mitigating the bounded recall problem. With the auditing agent $f$ effectively triggering resampling when necessary, we observe a consistent increase in candidate recall across all benchmarks, raising the average ranking upper bound by +3.92 pp (78.80$\rightarrow$82.72). This indicates that the agent successfully identifies scenarios where the initial candidate pool lacks the correct SQL.

This improvement is orthogonal to the ranking process: rankers select the best SQL within a fixed set, while the resampling module expands the search space to include correct candidates. Together, they ensure the ranker operates on a higher-quality candidate pool.

\begin{table}[t]
    \centering
    \small
    \begin{tabular}{l c}
        \toprule
        Method & EX (\%) \\
        \midrule
        \ours & \textbf{75.03} \\
        \phantom{0}\textit{w/o} \oursprm & 74.19 \\
        \bottomrule
    \end{tabular}
    \normalsize
    \caption{Position bias mitigation by \ours's pointwise ranker on BIRD-dev.}
    \label{tab:pointwise_ablation}
\end{table}
\begin{table*}[t]
    \centering
    \small
    \resizebox{\textwidth}{!}{
    \begin{tabular}{l c c c c}
        \toprule
        \makecell[l]{SQL Selection Method} & 
        \makecell{\# Pointwise Ranking Calls\\per Query} & 
        \makecell{\# Listwise Ranking Calls\\per Query} & 
        \makecell{Inference Time\\per Query (sec/query; $\downarrow$)} & 
        \makecell{EX (\%)} \\
        \midrule
        CSC-SQL & -- & -- & 0.00 & 71.58 \\
        Contextual-SQL & 32 & -- & 0.26 & 73.14 \\
        XiYan-SQL & -- & 2 & 0.03 & 72.03 \\
        CHASE-SQL & -- & 138 & 1.68 & 73.34 \\
        \ours & 32 & 107 & 1.56 & \textbf{75.03} \\
        \phantom{0}\textit{always-resample} & 32 & 138 & 1.96 & 74.25 \\
        \bottomrule
    \end{tabular}}
    \normalsize
    \caption{Computational overhead and performance trade-off of SQL selection methods on BIRD-dev. All inference times were measured on the same machine equipped with 8$\times$H200 GPUs.}
    \label{tab:computational_efficiency}
\end{table*}

\subsection{\textbf{Position Bias Mitigation}}
\label{analysis:position_bias}
To verify whether the position-consistency objective mitigates positional bias, we compare \ourslrm against two variants: without consistency reward and without GRPO training (Table~\ref{tab:consistency_reward}). The consistency reward yields the largest gain, improving input consistency by +11.89 pp over GRPO alone (45.60$\rightarrow$57.49). Table~\ref{tab:pointwise_ablation} further shows that removing the pointwise ranker drops EX by $-$0.84 pp (75.03$\rightarrow$74.19), confirming its role as an order-insensitive anchor.

\begin{table}[t]
    \centering
    \small
    \begin{tabular}{l c}
        \toprule
        Method & EX (\%) \\
        \midrule
        \ours & \textbf{75.03} \\
        \phantom{0}\textit{w/o Agentic Resampling} & 74.25 \\
        \phantom{00}\textit{w/o Pointwise Pruning} & 73.92 \\
        \phantom{000}\textit{w/o Exec. Group Scoring} & 73.47 \\
        \phantom{000}\textit{w/o Pointwise Ranker} & 73.34 \\
        \phantom{000}\textit{w/o Listwise Ranker} & 73.14 \\
        \bottomrule
    \end{tabular}
    \normalsize
    \caption{Ablation study of \ours on BIRD-dev. All variants rank SQL candidates generated by Arctic-Text2SQL-R1-32B ($T{=}0.8$).}
    \label{tab:ablation_sql_selection}
\end{table}

\subsection{\textbf{Computational Overhead}}
Table~\ref{tab:computational_efficiency} illustrates the trade-off between computational overhead and selection accuracy. Among the high-performing listwise approaches, CHASE-SQL serves as a strong baseline but incurs the highest computational cost (1.68 sec/query). In contrast, \ours achieves 75.03 EX while reducing inference time by 0.12 sec/query compared to CHASE-SQL.

This efficiency arises from ranking only distinct execution groups to avoid redundancy, and employing agentic resampling that activates for just 37.01\% of test instances. This targeted approach is 0.40 sec/query faster than the always-resample variant, while achieving even higher accuracy. 
\begin{table}[t]
    \centering
    \small
    \setlength{\tabcolsep}{5pt}
    \begin{tabular}{l l c}
        \toprule
        Resampling & Candidate Pool & EX (\%) \\
        \midrule
        \textit{None} & Original candidates ($S$) & 74.25 \\
        \textit{Always} & Resampled candidates ($\tilde{S}$) & 74.32 \\
        \midrule
        \multirow{3}{*}{\textit{Agentic}} 
          & Union ($S \cup \tilde{S}_{\text{top-}n}$) & 73.92 \\
          & Replace ($S \to \tilde{S}$) where $m{=}n$ & 74.05 \\
          & Replace ($S \to \tilde{S}$) where $m{>}n$ & \textbf{75.03} \\
        \bottomrule
    \end{tabular}
    \normalsize
    \caption{Effect of agentic resampling in \ours on BIRD-dev EX. All variants rank SQL candidates generated by Arctic-Text2SQL-R1-32B ($T{=}0.8$).}
    \label{tab:agentic_ablation}
\end{table}

        

\begin{table}[t]
    \centering
    \small
    \setlength{\tabcolsep}{6pt}
    \begin{tabular}{l c c c}
        \toprule
        Decision Class & Precision & Recall & F1-Score \\
        \midrule
        \textit{Trigger Resampling} & 93.27 & 56.02 & 70.00 \\
        \textit{Skip Resampling}    & 31.22 & 83.17 & 45.40 \\
        \bottomrule
    \end{tabular}
    \normalsize
    \caption{Precision, recall, and F1-score of Agent $f$ on resampling decisions. Candidates were generated by Arctic-Text2SQL-R1-32B ($T{=}0.8$).} 
    \label{tab:prf_results}
\end{table}


\subsection{Ablation Study}
\label{analysis:ablation}

To verify the contribution of the ranking and resampling modules in \ours, we conduct an ablation study on BIRD-dev. As shown in Table~\ref{tab:ablation_sql_selection}, every variant underperforms the full \ours system (75.03).
Specifically, removing agentic resampling causes a significant drop to 74.25 ($-$0.78~pp), confirming the importance of candidate coverage for bounded recall. Next, disabling pointwise pruning further reduces EX to 73.92 ($-$0.33~pp).

Based on this setup, we further ablate the ranking methodology. Removing execution-group scoring leads to a decline to 73.47 ($-$0.45~pp), demonstrating the importance of consistent rewards for execution-equivalent SQLs. Finally, using single rankers proves less effective than our dual-reward design: relying solely on the listwise ranker (\textit{w/o pointwise}) yields 73.34 ($-$0.58~pp), while relying only on the pointwise ranker (\textit{w/o listwise}) drops performance to 73.14 ($-$0.78~pp).

\subsection{Agentic Resampling}
\label{analysis:kimera_pool}
To validate the effectiveness of our agentic resampling module, we analyze both the downstream impact on execution accuracy (Table~\ref{tab:agentic_ablation}) and the agent's decision capability (Table~\ref{tab:prf_results}).

As shown in Table~\ref{tab:agentic_ablation}, naive strategies fail to yield significant gains. \textit{Always} resampling provides only a marginal improvement (+0.07~pp), while simply taking the \textit{Union} of original and resampled candidates actually degrades performance to 73.92, suggesting that indiscriminately adding candidates introduces noise that distracts the ranker. In contrast, our proposed approach selectively replaces the pool with a larger set ($m{>}n$) only when triggered, achieving the highest EX of 75.03.

The success of our method is explained by the agent's decision metrics in Table~\ref{tab:prf_results}. The agent demonstrates high precision (93.27) in triggering resampling, ensuring that valid candidate pools are rarely discarded (minimizing false positives). It targets only the most problematic queries for repair, and the high recall for the \textit{Skip} class (83.17) prevents unnecessary computational costs. 
\section{Conclusion}
In this paper, we introduced \ours, a Text-to-SQL framework that addresses functional inconsistency and bounded recall through unified reward modeling. \ours groups candidates by execution result and ranks groups using two complementary signals: a cross-group preference from pairwise comparisons and a single-group utility from pointwise scoring and group size. To address bounded recall, an agentic resampling module selectively expands the candidate pool when the correct SQL is likely absent. On BIRD-dev, \ours achieves 75.03 EX, a new state of the art among methods using models with disclosed sizes.

\section*{Acknowledgment}
This work was supported by the Institute of Information \& communications Technology Planning \& Evaluation (IITP) grant funded by the Korea government(MSIT) (No.2022-0-00995, Automated reliable source code generation from natural language descriptions) and ITRC(Information Technology Research Center) support program(IITP-2026-RS-2020-II201789) supervised by the IITP.

\section{Limitation}
While \ours excels in in-domain settings, the reliance on a supervised pointwise ranker constrains generalization. As observed in Tables~\ref{tab:cross_benchmark} and~\ref{tab:indomain_ood}, the domain gap in the pointwise ranker leads to a marginal performance difference (0.46--0.67) compared to the pointwise-ablated variant on out-of-domain benchmarks. Importantly, however, \ours consistently outperforms all baselines even with this module included. Integrating a domain-generalized pointwise ranker remains a promising direction for future work.


\bibliography{anthology,custom}

\setcounter{dbltopnumber}{1}
\renewcommand{\dbltopfraction}{0.7}
\appendix
\clearpage
\section*{\centering Appendices}

\begin{table*}[t]
\centering
\small
\setlength{\tabcolsep}{5pt}
\resizebox{\textwidth}{!}{
\begin{tabular}{llccccccc}
\toprule
\multirow{2.5}{*}{SQL Selection Method} &
\multirow{2.5}{*}{Ranking} &
\multicolumn{4}{c}{\textbf{In-domain}} &
\multicolumn{3}{c}{\textbf{Out-of-domain}} \\
\cmidrule(lr){3-6} \cmidrule(lr){7-9}
 &  & BIRD-dev & Spider-test & Spider-DK & \textbf{Avg.} & EHRSQL & ScienceBenchmark & \textbf{Avg.} \\
\midrule
\ours & Groupwise (Point + List) + FMV & \textbf{75.03} & \textbf{87.19} & \textbf{77.92} & \textbf{80.05} & 46.30 & 66.82 & 56.56 \\
\phantom{0}\textit{w/o \oursprm} & Listwise + FMV & 74.19 & 86.71 & 76.82 & 79.24 & \textbf{46.97} & \textbf{67.28} & \textbf{57.13} \\
\bottomrule
\end{tabular}}
\normalsize
\caption{
Performance on in-domain (BIRD-dev, Spider-test, Spider-DK) and out-of-domain (EHRSQL, ScienceBenchmark) datasets w.r.t. the ranker's training distribution. 
}
\label{tab:indomain_ood}
\end{table*}

\section{Further Analysis}
\subsection{Domain Generalization}
\label{appendix:domain_gen}
\begin{table*}[t]
    \centering
    \scriptsize
    \setlength{\tabcolsep}{6pt}
    \begin{tabular}{
        l 
        l 
        c 
        c 
        c 
        c 
        c 
        c
    }
        \toprule
        \multirow{2}{*}{SQL Selection Method} &
        \multirow{2}{*}{Ranking} &
        \multirow{2}{*}{BIRD-dev} &
        \multirow{2}{*}{Spider-test} &
        \multirow{2}{*}{Spider-DK} &
        \multirow{2}{*}{EHRSQL} &
        \multirow{2}{*}{\shortstack{Science\\Benchmark}} &
        \multirow{2}{*}{{Avg.}} \\
        \\
        \midrule
        \textit{Contextual-SQL} & & & & & & & \\
        ~~ \textit{w/} Contextual-RM-32B & Pointwise &
        73.01 &
        {86.54} &
        75.11 &
        41.01 &
        {63.36} &
        67.81 \\
        ~~ \textit{w/} \oursprm & Pointwise &
        {73.14} &
        86.36 &
        {75.50} &
        {41.41} &
        63.13 &
        {67.91} \\
        \midrule
        \textit{\ours} & & & & & & & \\
        ~~ \textit{w/} Contextual-RM-32B & Groupwise (Point + List) + FMV &
        74.90 &
        {87.29} &
        77.70 &
        45.79 &
        {66.82} &
        70.50 \\
        ~~ \textit{w/} \oursprm & Groupwise (Point + List) + FMV &
        {75.03} &
        87.19 &
        {77.92} &
        {46.30} &
        {66.82} &
        {70.65} \\
        \bottomrule
    \end{tabular}
    \normalsize
    \caption{
        EX comparison between Contextual-RM-32B and our optimized
        pointwise ranker, \oursprm, across five benchmarks.
    }
    \label{tab:pointwise_results}
\end{table*}

To validate the effectiveness of \ours across in-domain and out-of-domain (OOD), we conduct the experiments shown in Table~\ref{tab:indomain_ood}.
The \textit{w/o \oursprm} applies our groupwise selection while treating all candidates as having equal pointwise ranker scores.

As shown in Table~\ref{tab:indomain_ood}, \ours achieves slightly higher gains on OOD benchmarks when the pointwise ranker is excluded. We attribute this to the difference in training data diversity between the two ranking components. Our listwise ranker, \ourslrm, is based on OmniSQL-7B, which is trained on a broad corpus including BIRD, Spider, and SynSQL-2.5M~\cite{li2025omnisql}, which covers a wide spectrum of domains. In contrast, our pointwise ranker \oursprm is based on Contextual-RM-32B, which is fine-tuned exclusively on BIRD. Consequently, the pointwise scores are less reliable on unseen domains, creating a bottleneck that marginally offsets the generalized performance of the listwise ranker. Integrating a domain-generalized pointwise ranker in the future would further improve OOD performance.

\subsection{Binary Accuracy of Pairwise Rankers}
\begin{table}[h]
    \centering
    \scriptsize
    \begin{tabular}{l l c}
        \toprule
        Method & Base Model & Binary Acc. (\%) \\
        \midrule
        GRPO + \textit{consistency reward} & OmniSQL-7B & \textbf{78.86} \\                
        GRPO & OmniSQL-7B & 76.08 \\
        SFT & OmniSQL-7B & 67.66 \\
        \midrule
        GRPO + \textit{consistency reward} & Qwen2.5-Coder-7B & \textbf{76.85} \\        
        GRPO & Qwen2.5-Coder-7B & 74.22 \\
        \bottomrule
    \end{tabular}
    \normalsize
    \caption{Binary selection accuracy of pairwise rankers on BIRD-dev. 
GRPO + \textit{consistency reward} is our proposed ranker, while SFT represents the supervised selector used in approaches such as CHASE-SQL. 
Positive–negative pairs were generated using Qwen2.5-Coder-7B ($T{=}0.8$).}
    \label{tab:RM_acc}
\end{table}

To validate \ourslrm (\S\ref{sec:pairwise_rm}), which operates as a listwise ranker over candidate pairs, we report binary selection accuracy on BIRD-dev (Table~\ref{tab:RM_acc}). 
We compare our final ranker (GRPO + \textit{consistency reward}) against both GRPO and an SFT-based selector, which represents the selection module used in approaches such as CHASE-SQL. 
\ourslrm consistently outperforms the strongest baseline, GRPO, across different base models. Specifically, it achieves 78.86 on OmniSQL-7B (+2.78 pp over GRPO) and 76.85 on Qwen2.5-Coder-7B (+2.63 pp over GRPO), improving selection quality across both backbones.
Table~\ref{tab:RM_ex} further supports the superiority of \ourslrm on the end-to-end EX performance.

\subsection{Performance across $\tau$}
\label{appendix:tau}
\begin{figure}[t]
    \centering
    \includegraphics[width=1.0\linewidth]{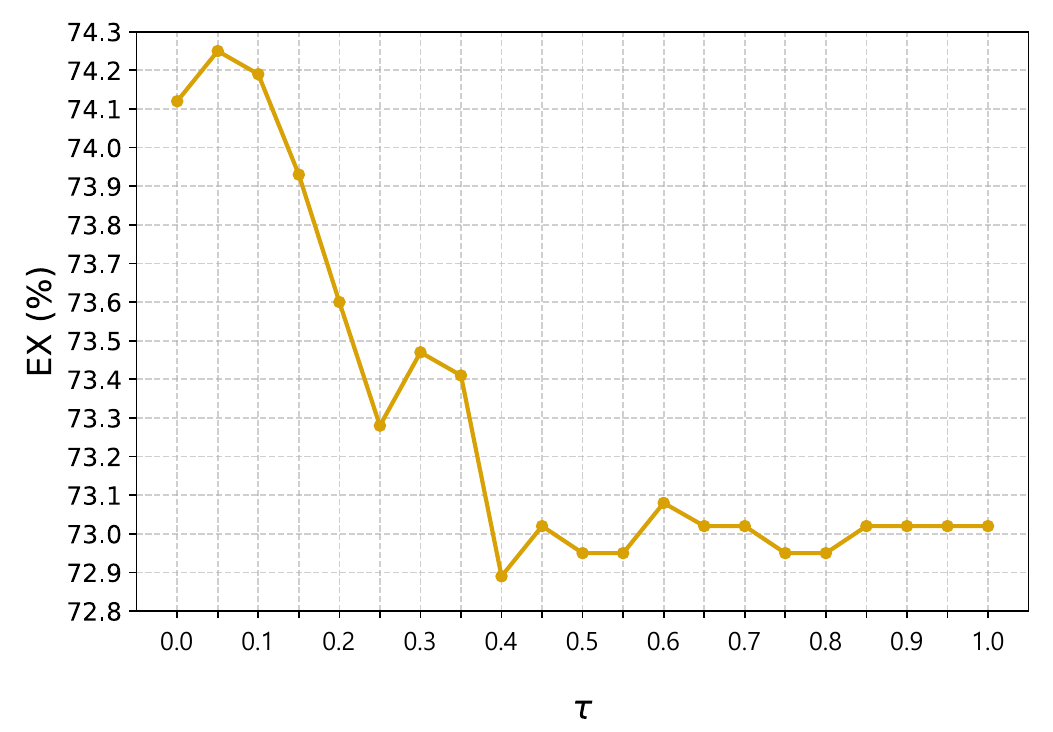}
    \caption{Execution accuracy (EX) of \ours on BIRD-dev across the threshold $\tau$ in Eq.~\eqref{eq:threshold}. Each point ranks $n{=}32$ SQL candidates per query generated by Arctic-Text2SQL-R1-32B with nucleus sampling ($T{=}0.8$).}
    \label{fig:tau}
\end{figure}

To evaluate how the threshold $\tau$ in Eq.~\eqref{eq:threshold} affects \ours's execution accuracy, Figure~\ref{fig:tau} shows execution accuracy as $\tau$ is varied. The curve peaks at very low $\tau$ and then declines as $\tau$ increases. At low $\tau$, \ours applies the normalized $[0,1]$ group-mean score only when the pairwise ranker strongly indicates that the group consists of incorrect candidates; otherwise the group is assigned the default score of 1. This confidence-gated strategy reduces bias from negative--negative pairings that can accidentally elevate incorrect candidate groups over correct ones, which is a practical source of false positives. Overall, the results favor tuning $\tau$ toward lower, more selective values so that listwise ranker is applied only when reliable, rather than always trusting fine-grained listwise scores.

\subsection{\textbf{Generalization to SQL Generators}}
To verify the generator-agnostic robustness of \ours, we replicate the evaluation using OmniSQL-7B (Table~\ref{tab:omnisql_results}). This smaller model creates a more challenging selection environment due to a lower yield of correct candidates. Despite the noisier candidate pool, \ours achieves 71.83 EX, outperforming the strongest baseline (Contextual-SQL, 70.80) by +1.03 pp. This confirms that our framework remains effective even with weaker underlying generators.

\subsection{\textbf{Generalization to Agentic Workflows}}
While our main experiments on 5 text2sql benchmarks already evaluate the core reasoning ability of SQL candidate selection, we conduct an additional evaluation on the Spider2.0~\cite{lei2025spider} to examine whether the benefit of \ours transfers to a more realistic setting. 
Following the evaluation protocol of~\citet{yao2025arctic}, we focus on the SQLite subset so that the comparison remains centered on the SQL generation/selection stage. 

As shown in Table~\ref{tab:spider2lite_sqlite}, \ours achieves the best EX score of 29.17, outperforming all baselines. Although this experiment is not a full end-to-end evaluation of an agentic system, it provides complementary evidence that our method remains effective under more realistic schema settings and can be naturally incorporated as a ranking and selection module within broader agentic Text-to-SQL pipelines.

\subsection{Efficiency Evaluation}
\label{appendix:rves}
To complement EX, we additionally evaluate selection methods using Reward-based Valid Efficiency Score (R-VES)\footnote{\href{https://github.com/bird-bench/mini\_dev}{https://github.com/bird-bench/mini\_dev}} on BIRD-dev, where R-VES extends execution-based evaluation by considering the efficiency of valid SQL queries. 

Table~\ref{tab:rves_bird_dev} reports the EX and R-VES scores of different selection methods.
\ours achieves the best performance on both metrics, obtaining 75.03 EX and 69.73 R-VES.
In particular, it outperforms the strongest baseline, CHASE-SQL, by 1.69 pp in EX and 1.04 pp in R-VES.
These results indicate that the advantage of \ours is not limited to execution correctness, but also extends to the efficiency of selected valid SQLs.

\subsection{\textbf{Robustness to Grouping Edge Cases}}
We analyzed potential failure modes of execution-based grouping: cases where all execution results are distinct (making grouping impossible) or all are empty (making grouping uninformative). Our empirical analysis on BIRD-dev reveals that the \textit{``all-distinct''} scenario is virtually non-existent (0\%), reflecting the inherent convergence of LLM outputs. Conversely, the \textit{``all-empty''} scenario occurred in 1.43\% (22/1,534) of queries. Critically, our framework does not fail in these instances; instead, the agentic resampling module detects the low confidence or absence of valid content and triggers the generation of new candidates to resolve the ambiguity.

\section{Discussion}
\subsection{\textbf{Robustness against Coincidental Correctness}}
A potential challenge in execution-based grouping is the occurrence of false positives, which are semantically incorrect SQLs that coincidentally yield the correct execution result. In our framework, these candidates are naturally grouped with the correct SQLs since they share the same execution output. While this introduces semantic noise within the target group, our scoring and selection mechanism is explicitly designed to filter out such \textit{``lucky guesses.''}
\begin{table}[t]
    \centering
    \scriptsize
    \begin{tabular}{l l c}
        \toprule
        Method & Base Model & EX (\%) \\
        \midrule
        GRPO + \textit{consistency reward} & OmniSQL-7B & \textbf{71.45} \\                
        GRPO & OmniSQL-7B & 70.79 \\
        SFT & OmniSQL-7B & 69.82 \\
        \midrule
        GRPO + \textit{consistency reward} & Qwen2.5-Coder-7B & \textbf{71.19} \\        
        GRPO & Qwen2.5-Coder-7B & 69.88 \\
        \bottomrule
    \end{tabular}
    \normalsize
    \caption{EX of pairwise rankers on BIRD-dev. 
GRPO + \textit{consistency reward} denotes our ranker, and SFT represents the supervised selector used in approaches such as CHASE-SQL. 
Positive–negative pairs were generated using Qwen2.5-Coder-7B ($T{=}0.8$).}
    \label{tab:RM_ex}
\end{table}

\begin{table}[t]
    \centering
    \small
    \begin{tabular}{
            l 
            c 
    }
        \toprule
        SQL Selection Method & BIRD-dev \\
        \midrule
        CSC-SQL & 67.40 \\
        Contextual-SQL & 70.80 \\
        CHASE-SQL & 68.38 \\
        XiYan-SQL & 68.06 \\
        \ours & \textbf{71.83} \\
        \bottomrule
    \end{tabular}
    \caption{EX comparison of different SQL selection methods on the BIRD-dev benchmark, where SQL candidates were generated by OmniSQL-7B.}
    \label{tab:omnisql_results}
\end{table}

\begin{table}[t]
\centering
\small
\begin{tabular}{l c}
\toprule
Method & Spider2.0-SQLite \\
\midrule
CSC-SQL & 12.50 \\
Contextual-SQL & 16.67 \\
CHASE-SQL & 20.83 \\
XiYan-SQL & 16.67 \\
\ours & \textbf{29.17} \\
\bottomrule
\end{tabular}
\caption{EX comparison of different selection methods on the Spider 2.0-SQLite, following~\citet{yao2025arctic}.}
\label{tab:spider2lite_sqlite}
\end{table}
\begin{table}[t]
\centering
\small
\begin{tabular}{l c c}
\toprule
Method & EX (\%) & R-VES \\
\midrule
CSC-SQL          & 71.58 & 67.29 \\
Contextual-SQL   & 73.14 & 67.80 \\
CHASE-SQL        & 73.34 & 68.69 \\
XiYan-SQL        & 72.03 & 67.28 \\
\ours & \textbf{75.03} & \textbf{69.73} \\
\bottomrule
\end{tabular}
\caption{Comparison of different selection methods on BIRD-dev using EX and Reward-based Valid Efficiency Score (R-VES). EX results are copied from Table~\ref{tab:cross_benchmark}. Higher is better for both metrics.}
\label{tab:rves_bird_dev}
\end{table}

The final representative for each group is determined by the candidate with the highest pointwise score (Eq~\eqref{eq:anchor}), not selected randomly. Since the pointwise ranker evaluates the semantic alignment between the SQL and the question, it assigns lower scores to false positives compared to valid SQLs. Consequently, even if false positives exist within the correct group, the group anchor remains a semantically correct SQL, ensuring that the final decision is guided by the most reliable candidate rather than by noise.

\subsection{\textbf{Case Studies: Functional Inconsistency in Real BIRD Failures}}
\label{app:token_bias_cases}
Figure~\ref{fig:token_bias_cases} presents two real BIRD-dev case studies, sorted by pointwise RM score. In both examples, multiple incorrect SQLs share the same execution result but receive different pointwise scores, revealing that pointwise RM is sensitive to token-level variations even when execution semantics are identical. 
As a result, the highest-scored candidate under the pointwise RM is incorrect in both cases. \ours resolves this issue by first grouping candidates by execution result and then applying the listwise RM over execution groups, which promotes the correct execution group to Top~1.

\section{Implementation Details}
\label{appendix:details}
\paragraph{Candidate Sampling.}
We generate the initial pool of SQL candidates using Arctic-Text2SQL-R1-32B~\cite{yao2025arctic}.
Following its guideline,\footnote{\href{https://github.com/snowflakedb/ArcticTraining/tree/main/projects/arctic_text2sql_r1}{Arctic-Text2SQL-R1's github}} we utilize the prompt template shown in Figure~\ref{fig:initial_pool_generate} and set the sampling temperature to 0.8.
For each question, we generate $n{=}32$ candidate SQLs in the initial pool, and selectively resample $m{=}1,024$ candidates.

\paragraph{Agentic Resampling.}
For the agentic resampling, agent $f$ must determine whether the initial candidate pool contains a correct answer.
Specifically, the agent observes the question, database schema, candidate SQL queries, and their execution results with the prompt shown in Figures~\ref{fig:agent_system_prompt} and~\ref{fig:agent_user_prompt}.
To effectively follow this instruction, training data is required, so we generate candidates from BIRD-train under same candidate generation setting as used at evaluation time.
We then construct examples where the pool does or does not contain the correct answer.
We evaluate three approaches using Qwen2.5-Coder-7B-Instruct \cite{hui2024qwen25coder} as the backbone: In-Context Learning (ICL)\footnote{ICL randomly samples few-shot from the training data.}, Supervised Fine-Tuning (SFT)\footnote{Trained with gold labels indicating answer presence.}, and GRPO\footnote{Reward of 1 for correct prediction, 0 otherwise.}.
Then, ICL performs worse (74.44), while SFT and GRPO achieve identical performance (75.03).
We adopt SFT, due to its efficiency. 
We use this LLM agent as the auditing agent $f$.

\paragraph{Rankers.}
To train our listwise ranker, \ourslrm, we use OmniSQL-7B~\cite{li2025omnisql} as the backbone model, which is based on Qwen2.5-Coder-7B-Instruct.
We construct training data by generating 32 SQL candidates from BIRD-train and Spider-train, then sampling one correct and one incorrect candidate per question to form positive-negative pairs.
Since the resampled pool contains many SQL candidates with high pointwise ranker scores, we sample hard negatives from the Top 15 highest-scoring incorrect candidates during training.
The prompt template for pairwise comparison is shown in Figure~\ref{fig:pairwise_rm_prompt}.
The sequence length is set to 8K tokens. 
When prompts exceed this limit, we proportionally truncate the execution results from both candidates to minimize information loss.

Our pointwise ranker, \oursprm, is initialized from Contextual-RM-32B.\footnote{\href{https://huggingface.co/ContextualAI/ctx-bird-reward-250121}{Contextual-RM-32B's huggingface}}
According to Contextual-SQL~\cite{agrawal2025text2sql}, Contextual-RM-32B is trained with Qwen2.5-Coder-32B-Instruct as its backbone.
We further optimize this model on BIRD-train by selecting hard negative SQL candidates sorted by \ourslrm and training with the same recipe as Contextual-SQL.
As shown in Table~\ref{tab:pointwise_results}, the additional training on BIRD-train show a slight gain on BIRD-dev, while no consistent trend is observed on other benchmarks.

\paragraph{Experimental Infrastructure.} 
All training and evaluation experiments are conducted on a single node equipped with 8 NVIDIA H200 GPUs.

\begin{table}[t]
    \centering
    \scriptsize
    \resizebox{1.0\linewidth}{!}{
    \begin{tabular}{lccc}
        \toprule
        Benchmark & \#Questions & \#Domains & Domain Type \\
        \midrule
        BIRD-dev & 1,534 & 37+ & Cross-domain \\
        Spider-test & 2,147 & 138 & Cross-domain \\
        Spider-DK & 535 & 138 & Cross-domain \\
        EHRSQL & 1,008 & 1 & Clinical \\
        ScienceBenchmark & 299 & 3 & Scientific \\
        \bottomrule
    \end{tabular}
    }
    \normalsize
    \caption{Statistics of evaluation splits across five Text-to-SQL benchmarks.}
    \label{tab:benchmark_statistics}
\end{table}

\section{Prompt Templates}
\label{appendix:guidance}

We provide the complete prompt templates used throughout our framework.
Figure~\ref{fig:initial_pool_generate} shows the initial SQL generation prompt with chain-of-thought reasoning. 
Figure~\ref{fig:pairwise_rm_prompt} presents the pairwise comparison prompt for training and evaluating \ourslrm. 
Figures~\ref{fig:agent_system_prompt} and~\ref{fig:agent_user_prompt} describe the system and user prompts for the agent-guided generation module, which audits candidate pools and produces structured guidance for resampling.

\begin{figure*}[t]
    \footnotesize
    \centering

    \begin{tcolorbox}[
        title=\textbf{Case 1: BIRD-dev Question 196},
        left=3pt, right=3pt, top=3pt, bottom=3pt,
        enhanced, breakable,
        coltext=black
    ]
    \textbf{Question.} In the non-carcinogenic molecules, how many contain chlorine atoms? \\[0.5ex]

    \textbf{SQL1} (\emph{Incorrect}; pointwise RM score: $-3.8625$) \\
    {\ttfamily
    SELECT COUNT(T2.molecule\_id) \\
    FROM molecule AS T1 \\
    JOIN atom AS T2 ON T1.molecule\_id = T2.molecule\_id \\
    WHERE T1.label = `-' AND T2.element = `cl';
    \par}
    Execution result: \texttt{[[201]]} \\[0.5ex]

    \textbf{SQL2} (\emph{Correct}; pointwise RM score: $-4.8750$) \\
    {\ttfamily
    SELECT COUNT(DISTINCT m.molecule\_id) \\
    FROM molecule AS m \\
    JOIN atom AS a ON m.molecule\_id = a.molecule\_id \\
    WHERE m.label = `-' AND a.element = `cl';
    \par}
    Execution result: \texttt{[[71]]} \\[0.5ex]

    \textbf{SQL3} (\emph{Incorrect}; pointwise RM score: $-5.0250$) \\
    {\ttfamily
    SELECT COUNT(*) \\
    FROM atom \\
    WHERE molecule\_id IN ( \\
    \hspace*{1.5em}SELECT molecule\_id \\
    \hspace*{1.5em}FROM molecule \\
    \hspace*{1.5em}WHERE label = `-' \\
    ) AND element = `cl';
    \par}
    Execution result: \texttt{[[201]]} \\[0.5ex]

    \textbf{Observation.} SQL1 and SQL3 count chlorine atoms rather than distinct molecules containing chlorine. Pointwise RM ranks \textbf{SQL1 $>$ SQL2 $>$ SQL3}, although SQL1 and SQL3 belong to the same incorrect execution group (\texttt{[[201]]}). R$^3$-SQL instead ranks the correct execution group (\textbf{SQL2}, \texttt{[[71]]}) as \textbf{Top~1} and the incorrect execution group (\textbf{SQL1, SQL3}, \texttt{[[201]]}) as \textbf{Top~2}.
    \end{tcolorbox}

    \vspace{0.5em}

    \begin{tcolorbox}[
        title=\textbf{Case 2: BIRD-dev Question 1527},
        left=3pt, right=3pt, top=3pt, bottom=3pt,
        enhanced, breakable,
        coltext=black
    ]
    \textbf{Question.} Which gas station has the highest amount of revenue? \\[0.5ex]

    \textbf{SQL1} (\emph{Incorrect}; pointwise RM score: $-5.3625$) \\
    {\ttfamily
    SELECT GasStationID \\
    FROM transactions\_1k \\
    GROUP BY GasStationID \\
    ORDER BY SUM(Amount * Price) DESC \\
    LIMIT 1;
    \par}
    Execution result: \texttt{[[3250]]} \\[0.5ex]

    \textbf{SQL2} (\emph{Correct}; pointwise RM score: $-5.8125$) \\
    {\ttfamily
    SELECT GasStationID \\
    FROM transactions\_1k \\
    GROUP BY GasStationID \\
    ORDER BY SUM(Amount) DESC \\
    LIMIT 1;
    \par}
    Execution result: \texttt{[[4347]]} \\[0.5ex]

    \textbf{SQL3} (\emph{Incorrect}; pointwise RM score: $-6.1125$) \\
    {\ttfamily
    SELECT T2.GasStationID \\
    FROM transactions\_1k AS T1 \\
    INNER JOIN gasstations AS T2 ON T1.GasStationID = T2.GasStationID \\
    GROUP BY T2.GasStationID \\
    ORDER BY SUM(T1.Price * T1.Amount) DESC \\
    LIMIT 1;
    \par}
    Execution result: \texttt{[[3250]]} \\[0.5ex]

    \textbf{Observation.} SQL1 and SQL3 use \texttt{SUM(Amount * Price)}, while the correct formulation under the dataset semantics is \texttt{SUM(Amount)}. Pointwise RM ranks \textbf{SQL1 $>$ SQL2 $>$ SQL3}, although SQL1 and SQL3 belong to the same incorrect execution group (\texttt{[[3250]]}). R$^3$-SQL instead ranks the correct execution group (\textbf{SQL2}, \texttt{[[4347]]}) as \textbf{Top~1} and the incorrect execution group (\textbf{SQL1, SQL3}, \texttt{[[3250]]}) as \textbf{Top~2}.
    \end{tcolorbox}

    \caption{\edit{Real BIRD-dev case studies illustrating functional inconsistency in pointwise ranking, sorted by pointwise RM score (higher is better).} }
    \label{fig:token_bias_cases}
\end{figure*}
\begin{figure*}[t]
    \small
    \centering
    \begin{tcolorbox}[
        title=\textbf{Initial Pool Generation Prompt},
        left=3pt, right=3pt, top=3pt, bottom=3pt,
        enhanced, breakable
    ]
        \textbf{System:} \\
        You are a data science expert. Below, you are provided with a database schema and a natural language question. Your task is to understand the schema and generate a valid SQL query to answer the question. \\[1ex]
        
        \textbf{User:} \\
        Database Engine: \\
        SQLite \\[0.5ex]
        
        Database Schema: \\
        \texttt{\{Database Schema\}} \\
        This schema describes the database's structure, including tables, columns, primary keys, foreign keys, and any relevant relationships or constraints. \\[1ex]
        
        Question: \\
        \texttt{\{evidence + question\}} \\[1ex]
        
        Instructions: \\
        - Make sure you only output the information that is asked in the question. If the question asks for a specific column, make sure to only include that column in the SELECT clause, nothing more. \\
        - The generated query should return all of the information asked in the question without any missing or extra information. \\
        - Before generating the final SQL query, please think through the steps of how to write the query. \\[1ex]
        
        Output Format: \\
        Please provide a detailed chain-of-thought reasoning process and include your thought process within \verb|<thinking>| tags. Your final answer should be enclosed within \verb|<answer>| tags. Ensure that your SQL query follows the correct syntax and is formatted as follows: \\[0.5ex]
        
        \verb|```sql| \\
        \verb|-- Your SQL query here| \\
        \verb|```| \\[1ex]
        
        Example format: \\
        \verb|<thinking>| \\
        Step-by-step reasoning, including self-reflection and corrections if necessary. [Limited by 4K tokens] \\
        \verb|</thinking>| \\
        
        \verb|<answer>| \\
        Summary of the thought process leading to the final SQL query. [Limited by 1K tokens] \\[0.5ex]
        
        \verb|```sql| \\
        Correct SQL query here \\
        \verb|```| \\
        \verb|</answer>| \\
    \end{tcolorbox}
    \caption{Prompt for the Initial Pool Generation stage, where the LLM generates SQL candidates with chain-of-thought reasoning.}
    \label{fig:initial_pool_generate}
\end{figure*}
\begin{figure*}[t]
    \small
    \centering
    \begin{tcolorbox}[
        title=\textbf{Prompt Template for Training and Evaluating Pairwise Reward Model},
        left=3pt, right=3pt, top=3pt, bottom=3pt,
        enhanced, breakable
    ]

        \textbf{System:} \\
        You are a SQL expert. When given a SQL question along with two proposed solutions candidates A and B, your task is to evaluate the options based on clarity, efficiency, and adherence to best practices. Provide your answer strictly as either "A" or "B". \\[1ex]
        
        \textbf{User:} \\[0.5ex]
        Instruction: \\
        Given the DB info and question, there are two candidate queries. There is correct one and incorrect one. \\
        - First, think why one candidate is better than the other by comparing the two candidate answers and analyzing the differences of the query and the result. \\
        - Then, based on your analysis, the original question, and the provided database info, select the better candidate query. \\
        - Do not generate a new SQL query; focus solely on comparing the two given candidates. \\[0.5ex]
        ************************** \\[0.5ex]
        Database Schema \\
        \texttt{\{Database Schema\}}
        
        ************************** \\[0.5ex]
        Question: \\
        \texttt{\{Question\}}
        
        ************************** \\[0.5ex]
        Candidate A \\
        \texttt{\{SQL Query A\}} \\
        Execution result \\
        \texttt{\{Execution Result A\}}
        
        ************************** \\[0.5ex]
        Candidate B \\
        \texttt{\{SQL Query B\}} \\
        Execution result \\
        \texttt{\{Execution Result B\}}
        
        ************************** \\[0.5ex]
        Output Format: \\
        Please provide a detailed chain-of-thought reasoning process and include your thought process within \texttt{<think>} tags. Your final answer should be enclosed within \texttt{<answer>} tags.
        
        ************************** \\[0.5ex]
        Example format: \\
        \texttt{<think>} Step-by-step reasoning, including self-reflection and corrections if necessary. [Limited by 4K tokens] \texttt{</think>} \\
        \texttt{<answer>} Only write ``A'' or ``B'' depending on which is the correct answer. Do not include any other text. \texttt{</answer>} 
        
        \textbf{Assistant:} \\
        Let me solve this step by step. \\
        \texttt{<think>}
    \end{tcolorbox}
    \caption{Prompt template for pairwise reward model comparing two SQL candidates.}
    \label{fig:pairwise_rm_prompt}
\end{figure*}
\begin{figure*}[t]
    \small
    \centering
    \begin{tcolorbox}[
        title=\textbf{System Prompt in Agentic SQL Candidate Generation},
        left=3pt, right=3pt, top=3pt, bottom=3pt,
        enhanced, breakable
    ]

        You are a SQL Candidate Gatekeeper. \\
        Your job is to decide whether there is AT LEAST ONE LIKELY-CORRECT SQL among the given candidates. \\[1ex]

        Use the following criteria to judge whether a candidate is ``likely-correct'': \\
        1) Intent match: entities, filters, metrics, order, and top-k behavior align with the user query. \\
        2) Schema validity: the query uses correct tables/columns, required joins are present, and aggregations are legal. \\
        3) Execution sanity: the exec\_preview has a plausible shape/values for the query (no obvious contradictions). \\
        4) No major red flags: units/ratios are handled reasonably, limit/order are coherent, and there are no clearly spurious tables or conditions. \\[1ex]

        You should make a balanced judgment: \\
        - Mark \texttt{likely\_has\_correct=true} if at least one candidate appears reasonably correct according to the above criteria. \\
        - Minor ambiguities are acceptable as long as the query and SQL are broadly aligned and there are no obvious fatal issues. \\[1ex]

        Special handling for the first candidate (sorted pool): \\
        - Always inspect the first candidate carefully first. \\
        - If multiple candidates are likely-correct, prefer the first candidate as \texttt{best\_cand\_idx} when it also appears likely-correct. \\
        - If the first candidate is clearly incorrect, then consider other candidates for \texttt{best\_cand\_idx}. \\[1ex]

        You MUST output JSON with a SINGLE top-level key \texttt{"decision"}: \\[0.5ex]

        \texttt{\{} \\
        \phantom{xx}\texttt{"decision": \{} \\
        \phantom{xxxx}\texttt{"likely\_has\_correct": true/false,} \\
        \phantom{xxxx}\texttt{"confidence": 0.0\textasciitilde1.0,} \\
        \phantom{xxxx}\texttt{"reason\_tags": ["MISMATCH\_INTENT","MISSING\_JOIN", ...],} \\
        \phantom{xxxx}\texttt{"support": \{} \\
        \phantom{xxxxxx}\texttt{"best\_cand\_idx": <int or null>,} \\
        \phantom{xxxxxx}\texttt{"notes": "≤200 chars optional"} \\
        \phantom{xxxx}\texttt{\}} \\
        \phantom{xx}\texttt{\}} \\
        \texttt{\}} \\[1ex]

        Strict output rules: \\
        - Do NOT output any other top-level keys (NO \texttt{"sampling"}, NO \texttt{"guidance"}, NO \texttt{"compat\_drop\_mode"}). \\
        - Keep all fields in \texttt{"decision"} present; if unknown, use \texttt{null} or \texttt{[]} rather than omitting. \\
        - \texttt{cand\_idx} refers to ids provided with candidates (not array positions); use the given \texttt{cand\_idx} integers. \\
        - Output MUST be valid JSON, with double quotes on all keys and string values.

    \end{tcolorbox}
    \caption{Full system prompt for the LLM agent $f$ for resampling decision.}
    \label{fig:agent_system_prompt}
\end{figure*}

\begin{figure*}[t]
    \small
    \centering
    \begin{tcolorbox}[title=\textbf{User Prompt in Agentic SQL Candidate Generation}]
        You are given a Text2SQL problem. \\[0.5ex]
        
        \#\# User query \\
        \texttt{\{user\_query\}} \\[0.5ex]
        
        \#\# DB dialect \\
        \texttt{\{db\_dialect\}} \\[0.5ex]
        
        \#\# Schema (summary or DDL) \\
        \texttt{\{db\_schema\}} \\[0.5ex]
        
        \#\# Candidate SQLs (with id and execution preview) \\
        Each item provides (cand\_idx, SQL, and exec\_preview: a few rows from a dry-run; \texttt{None} means error/missing). \\
        The first candidate in the sorted pool is the primary candidate; pay particular attention to whether it is likely-correct. \\[0.5ex]
        
        \texttt{\{items\_block\}} \\[0.5ex]
        
        \#\#\# Matching checklist (for your internal reasoning) \\
        - Entities/filters/metrics/order/top-k extracted from user query must be reflected in the SQL. \\
        - If metric and filter come from different tables, valid explicit JOIN is required. \\
        - Ratios should reasonably avoid wrong integer division (e.g., using casting when appropriate). \\
        - Avoid undocumented mappings; rely only on provided schema/metadata. \\[0.5ex]
        
        Return JSON ONLY with a single top-level key \texttt{"decision"}, following the schema in the system prompt.
    \end{tcolorbox}
    \caption{Full input prompt for the LLM agent $f$ for resampling decision, 
    populated with the natural language question, schema, and candidate SQL pool.}
    \label{fig:agent_user_prompt}
\end{figure*}

\end{document}